# The polytropic behavior of Solar Wind protons as observed by Ulysses spacecraft during Solar minimum

[1]Georgios Nicolaou, [2]George Livadiotis, [2]David J. McComas

[1]Department of Space and Climate Physics, Mullard Space Science Laboratory, University College London, Dorking, Surrey, RH5 6NT, UK

[2]Department of Astrophysical Sciences, Princeton University, 171 Broadmead, Princeton, NJ, 08540, US

**Abstract**

We analyze proton bulk parameters derived from Ulysses observations and investigate the polytropic behavior of solar wind protons over a wide range of heliocentric distances and latitudes. The large-scale variations of the proton density and temperature over heliocentric distance, indicate that plasma protons are governed by sub-adiabatic processes (polytropic index $\gamma < 5/3$), if we assume protons with three effective kinetic degrees of freedom. From the correlation between the small-scale variations of the plasma density and temperature in selected sub-intervals, we derive a polytropic index $\gamma \sim 1.4$ on average. Further examination shows that the polytropic index does not depend on the solar wind speed. This agrees with the results of previous analyses of solar wind protons at $\sim 1$ au. We find that the polytropic index varies slightly over the range of the heliocentric distances and heliographic latitudes explored by Ulysses. We also show that the homogeneity of the plasma and the accuracy of the polytropic model applied to the data-points vary over Ulysses orbit. We compare our results with the results of previous studies which derive the polytropic index of solar wind ions within the heliosphere using observations from various spacecraft. We finally discuss the implications of our findings in terms of heating mechanisms and the effective degrees of freedom of the plasma protons.

## 1. Introduction

The polytropic equation (e.g., Chandrasekhar 1967) describes the correlation between the density $n$ and the temperature $T$ (or pressure $P \propto nT$) of a fluid, through the polytropic index $\gamma$

$$T \propto n^{\gamma-1}. \qquad (1)$$

The polytropic equation becomes vital in space plasma studies as it brings closure to the hierarchy of the plasma moments (e.g., Kuhn et al., 2010). Additionally, the use of the polytropic relationship can simplify the studies of complicated mechanisms including energy transfer between particles and fields (e.g., Kartalev et al. 2006). More specifically, the value of $\gamma$ is indicative for the energy transfer involved in the observed process and the effective degrees of freedom $f$ of the plasma particles. For example, during an adiabatic process, there is no heat transfer during the plasma compression or expansion. In this case, the polytropic index is equal to the ratio of the specific heats $\frac{c_p}{c_v}$, and is related to the kinetic degrees of freedom of the plasma particles $f$, such as $\gamma = 1 + \frac{2}{f}$. For instance, $\gamma = 5/3$ corresponds to adiabatic plasma particles with three kinetic degrees of freedom ($f = 3$). In another special case, an isothermal process (constant $T$) is characterized by $\gamma = 1$. During this process, the energy supplied to the system as heat balances the energy supplied to the system as work. Additionally, $\gamma = 0$ corresponds to isobaric plasmas (constant $P$) and $\gamma = \infty$ to isohoric plasmas, in which $n$ is constant (for more see Livadiotis 2016,

2019; Nicolaou et al. 2020). Therefore, $\gamma$ is a useful tool we can use to understand the nature of physical plasma mechanisms without solving complicated energy equations (e.g., Bavassano et al. 1996). Interestingly, although space plasmas are generally only very weakly collisional, we can describe some of their aspects using the fluid description and the polytropic equation, over a wide range of timescales (Verscharen et al. 2019; Wu et al. 2019; Coburn et al. 2022).

Several studies apply the polytropic model to particle observations in planetary magnetospheres. For example, Arridge et al. 2009 fit a polytropic model to electron observations, obtained by Cassini in Saturn's magnetotail plasma sheet. Their analysis shows that the electrons in the specific region exhibit an isothermal behavior. Dialynas et al. 2018 use Cassini observations to derive the energetic ion moments in Saturn's magnetosphere. The authors then apply a polytropic model to the derived density and temperature, revealing different behaviors for the observed species within the explored region. Nicolaou et al., 2014a;2015, derive the bulk parameters of plasma protons in the deep Jovian magnetotail using observations by New Horizons and identify intervals with anticorrelated $n$ and $T$. Linear $\log_{10}(T)$ vs $\log_{10}(n)$ fits to these intervals determine an isobaric relationship, indicating that the spacecraft crossed streamlines that were in pressure balance, or streamlines with plasma with isobaric behavior ($\gamma$=0). In another study, Park et al. 2019 use THEMIS observations to identify intervals where the plasma $n$ and $T$ are correlated. The analysis of these intervals makes it possible to calculate the polytropic index of ions in the magnetosheath of Earth and finds that $\gamma$ varies with the bow-shock geometry.

Numerous other studies investigate the polytropic behavior of solar wind plasma in the inner heliosphere including at ~1 au. For instance, Totten et al. 1995 use Helios observations to derive empirical relationships describing the proton density and temperature as functions of the heliocentric distance. These relationships determine a sub-adiabatic behavior ($\gamma$ ~1.46) on average. In another study, Newbury et al, 1997 use Pioneer Venus Orbiter measurements in the vicinity of stream interaction regions. They show a strong correlation between the proton $n$ and $T$, which determines an adiabatic polytropic index $\gamma$ ~5/3, and occasionally $\gamma$ ~2, suggesting that the degrees of freedom may be restricted. Kartalev et al. 2006 evaluate the Bernoulli integral fluctuations in the solar wind in order to select suitable time intervals for further analysis to estimate $\gamma$ of solar wind protons and electrons at ~1au. The study by Nicolaou et al. 2014b uses a similar method to analyze proton observations obtained from multiple spacecrafts at ~1au, between 01/01/1995 and 30/06/2012. Their analysis derives an average $\gamma$ ~1.8. The authors discuss the variations of $\gamma$ as a function of the solar activity and possible biases due to the instrument providing the observations. Several analyses of Wind spacecraft observations employ a very similar method to characterize the polytropic behavior of solar wind protons at ~1 au (e.g., Livadiotis & Desai 2016; Livadiotis 2018a,b; Nicolaou & Livadiotis 2019; Livadiotis & Nicolaou 2021, Nicolaou et al., 2021a,b). These analyses calculate a nearly adiabatic plasma with $\gamma$ ~5/3 on average. Moreover, the results show that $\gamma$ does not depend on the plasma speed or the solar activity. Recently, Nicolaou et al., 2020 analyzed high time-resolution observations of Solar Wind protons by Parker Solar Probe, revealing large-scale variations characterized by $\gamma$~5/3 and short time-scale fluctuations with $\gamma$ ~2.7. The authors discuss possible mechanisms that restrict the effective degrees of freedom or involve energy transfer from/to the plasma.

The analysis of measurements in the outer heliosphere, offer unique opportunities to study the evolution of the solar wind behavior as it propagates away from the Sun. For example, Elliott et al., 2019 analyze solar wind proton observations by New Horizons, obtained in the heliocentric distance range spanning from 21 to 43 au. A linear fitting to $\log_{10}(T)$ and $\log_{10}(n)$ in selected time-subintervals calculates the variation of $\gamma$ over heliocentric distance. Those authors found that $\gamma$ approaches ~0 (isobaric process) in the outer heliosphere. Interestingly, analyses of Interstellar Boundary Explorer observations determine $\gamma \sim 0$ for the ion plasma in the inner heliosheath (e.g., Livadiotis et al., 2011, 2013; Livadiotis & McComas 2013).

A complete understanding of the solar wind polytropic behavior within the entire heliosphere requires the analysis of measurements obtained across multiple locations throughout the system. In this study we analyze Ulysses observations which are obtained over a range of heliocentric distances and importantly, almost all heliographic latitudes. We explore if the plasma is described with a polytropic relationship and if this relationship depends on the plasma speed, heliocentric distance, and/or heliographic latitude. In Section 2, we describe the proton parameters we use for our analyses. In Section 3, we describe our methodology. In Section 4, we show our results which we discuss extensively in Section 5.

## 2. Data

Figure 1 shows the spacecraft attitude information and plasma parameters we analyze in this study. Among the spacecraft heliocentric distance $R$ and heliographic latitude $\Theta$, we use the 4 min resolution proton bulk parameters which are derived by fitting SWOOPS observations (Bame et al. 1992) to bi-Maxwellian core-beam distributions. We use only the core proton (slow and dense population) parameters. The details for the fit parameters are found at spase://NASA/NumericalData/Ulysses/SWOOPS/Proton/FitParameters/PT4M. For our analysis, we specifically use the proton core density $n$, scalar temperature $T$, flow speed $V$, and Alfvén speed $V_A = \frac{B}{\sqrt{\mu_0 \rho}}$, where $B$ is the magnetic field, $\rho$ is the mass density and $\mu_0$ the permeability constant. The scalar temperature $T$ is calculated from the thermal speeds parallel and perpendicular to the magnetic field $V_{th,\parallel}$ and $V_{th,\perp}$, respectively, such as $T = \frac{1}{3}\left[\frac{1}{2}m_p V_{th,\parallel}^2 + m_p V_{th,\perp}^2\right]$, where $m_p$ is the mass of the proton. We analyze the time-period spanning from 01/01/1992 to 31/12/1998, which is a period of a minimum solar activity, during which the solar wind is well structured as a function of the heliographic latitude (see McComas et al., 1998; 2008). As shown in Figure 1, during the analyzed time-period, most of the fast solar wind ($V > 600$ kms$^{-1}$) is observed in high absolute heliographic latitudes ($|\Theta| > 40°$).

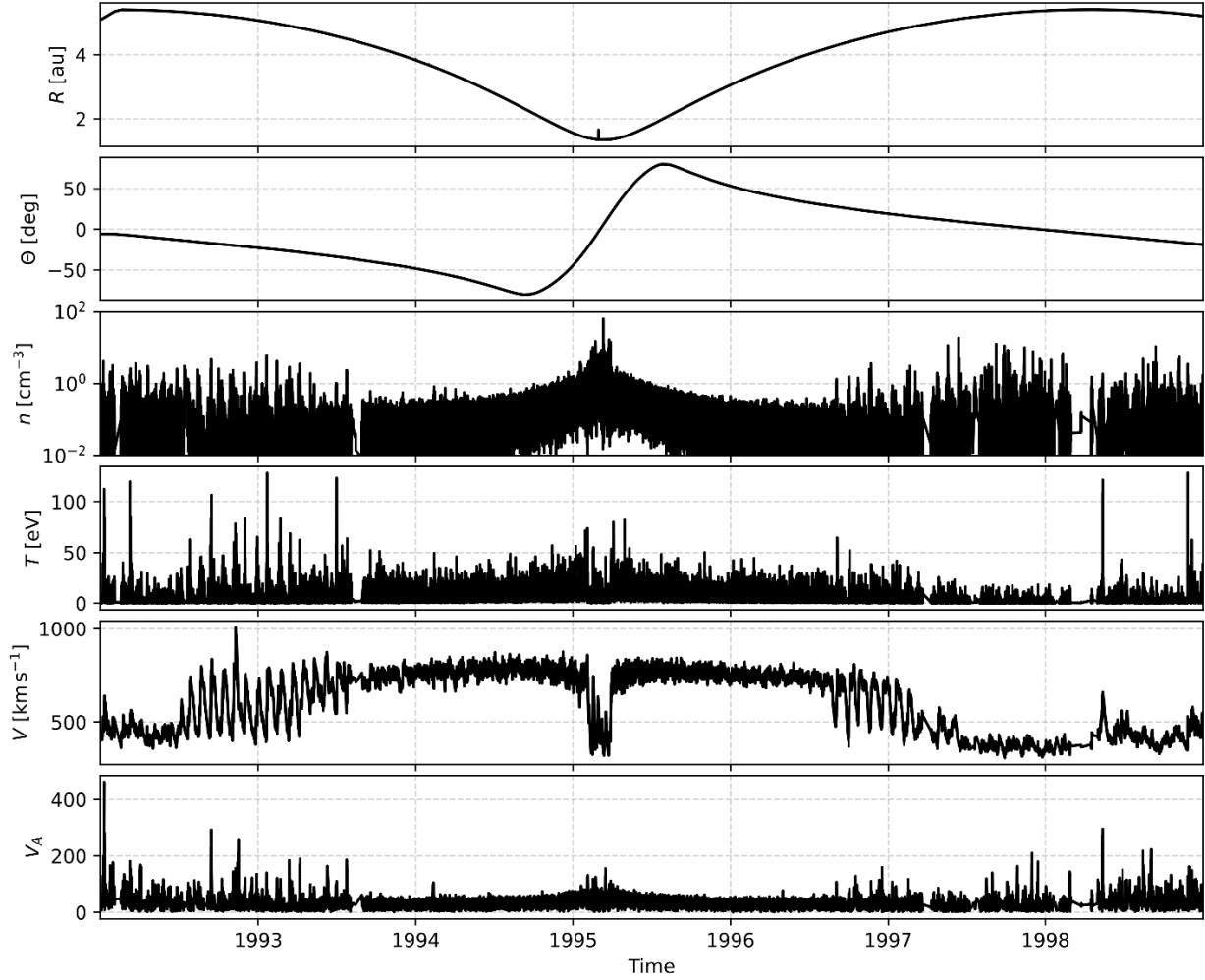

**Figure 1.** Time series of parameters we use in our analysis. From top to bottom we show the heliocentric distance of the spacecraft $R$, its heliographic latitude $\Theta$, the plasma core density $n$, temperature $T$, flow bulk speed $V$, and Alfvén speed $V_\mathrm{A}$.

## 3. Methodology

### 3.1 Histograms of large-scale variations

Similar to Nicolaou et al. 2020, we first examine the large-scale variations of the plasma density and temperature as functions of the heliocentric distance by constructing the 2D histograms of $\log_{10}(n)$ vs $\log_{10}(R)$ and $\log_{10}(T)$ vs $\log_{10}(R)$, respectively. The resolution we use for these histograms is $\Delta\log_{10}(n/\mathrm{cm}^{-3})\times \Delta\log_{10}(R/\mathrm{au}) = 0.075\times 0.01$ and $\Delta\log_{10}(T/\mathrm{eV})\times \Delta\log_{10}(R/\mathrm{au}) = 0.05\times 0.01$, respectively. Further, we normalize each column of these histograms to the maximum occurrence recorded within the specific $\log_{10}(R)$ bin. As explained in Livadiotis & Desai 2016, and later on in Livadiotis 2018b and Nicolaou et al. 2020, when there are not extreme values in the sampled dataset this normalization brings all the columns to the same level of occurrence and the correlation between the examined parameters becomes clearer. We then examine the correlation between the large-scale variations of the plasma density and temperature by constructing the 2D histogram of $\log_{10}(T)$ vs $\log_{10}(n)$ with resolution $\Delta\log_{10}(T/\text{eV})\times \Delta\log_{10}(n/\text{cm}^{-3})= 0.05\times0.075$. Again, since there are no extreme values in our samples, in order to reveal the correlation between the two parameters, we normalize the occurrence values in each $\log_{10}(T)$ bin (each column) to the maximum occurrence recorded within the bin.

**3.2 Linear fits to short-scale variations**

To investigate the short time-scale variations of $n$ and $T$ and identify their correlation, we choose sub-intervals of eight, consecutive data-points in our time series. Due to non-uniform data gabs in the data, the duration of the selected intervals varies. In Figure 2, we show the histogram of the duration of the subintervals we analyze in this study. The longest sub-interval we analyze does not exceed 200 minutes.

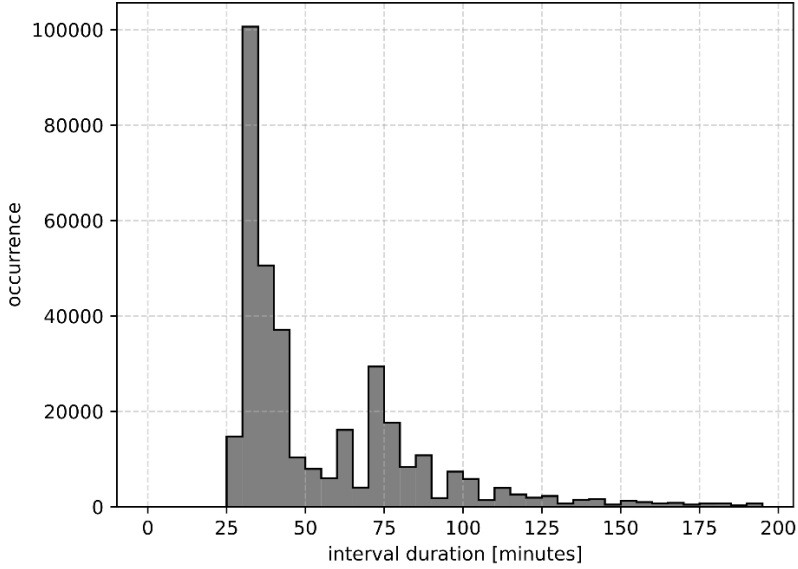

**Figure 2.** Histogram of the analyzed subinterval length.

Following a similar methodology to other studies deriving the polytropic index from single spacecraft observations, we perform a linear regression to $\log_{10}(T)$ vs $\log_{10}(n)$ within each sub-interval (e.g., Newbury et al. 1997; Arridge et al. 2009; Livadiotis 2018a,b; Nicolaou et al. 2021a,b,c). The slope of the fitted lines determines $\gamma$ since, according to Eq. 1:

$$\log_{10}(T) = (\gamma - 1) \log_{10}(n) + \text{const.} \qquad (2)$$

For each subinterval, we calculate the Pearson correlation coefficient of $\log_{10}(T)$ vs $\log_{10}(n)$ and the residuals between the data-points and the fitted line. We use these parameters to evaluate the applicability of the polytropic model to the observations. We also calculate the standard deviation and the mean value of the Bernoulli integral in each subinterval. We use the Bernoulli integral (*BI*) for an adiabatic plasma ($\gamma=5/3$):

$$BI = \tfrac{1}{2}V^2 + \tfrac{5k_B T}{2m} + \tfrac{1}{2}V_A^2, \qquad (3)$$

which is a reasonable simplification in our case as we discuss in the Appendix. The ratio of the standard deviation of the Bernoulli integral over its mean value $\frac{\text{std}(BI)}{\text{mean}(BI)}$ within each analyzed subinterval is proportional to the homogeneity of the analyzed plasma and indicates whether the data-points we analyze correspond to the same streamline or not (e.g., Kartalev et al. 2006; Nicolaou et al. 2014b; Pang et al. 2015; Livadiotis 2018a,b; Nicolaou 2021a,b). In Figure 3, we show three examples of such subintervals; one with $\gamma$ ~5/3, another with $\gamma$ ~2, and another with $\gamma$ ~1.2. All subintervals shown in Figure 3, exhibit a strong linear correlation (Pearson coefficient > 0.9) between $\log_{10}(T)$ and $\log_{10}(n)$ and they have a $\frac{\text{std}(BI)}{\text{mean}(BI)}$ ratio < 10%.

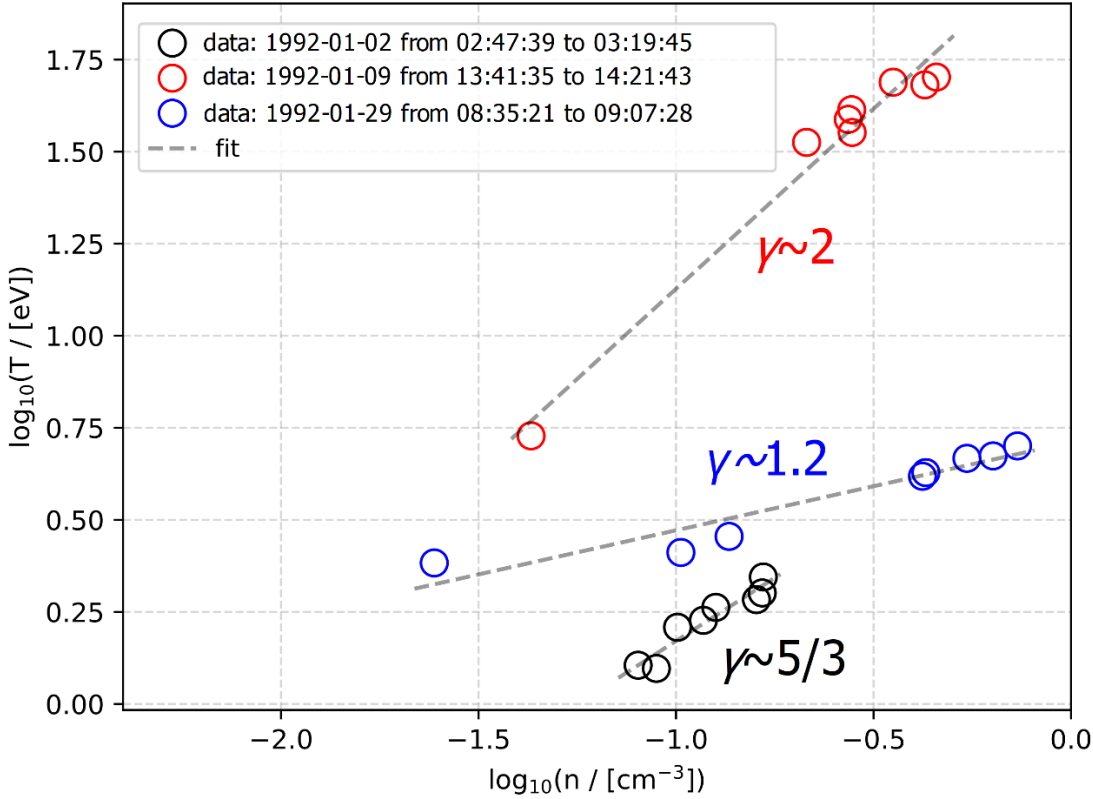

**Figure 3.** Three sub-intervals of $\log_{10}(T)$ vs $\log_{10}(n)$. Each set of colorful circles are data-points within subintervals exhibiting a strong correlation between $\log_{10}(T)$ vs $\log_{10}(n)$ (Pearson coefficient > 0.9), while the grey dashed lines are fitted linear models to the data. Black data-points correspond to an interval that is described with $\gamma$ ~ 5/3, while red data-points are fitted with $\gamma$ ~2, and blue data-points with $\gamma$ ~1.2.

## 4. Results

### 4.1. Large time-scale variations

In the top-left panel of Figure 4, we show the 2D histogram of $\log_{10}(n)$ vs $\log_{10}(R)$. The magenta lines on the same panel show the relationship $n \propto R^{-2}$ which describes the spherical expansion of steady speed plasma. Our result does not show any apparent deviation from the spherical expansion model for 1.38 au < $R$ < 4 au (0.14 < $\log_{10}(R/\text{au})$ < 0.6). For $R$ < 1.38 au and $R$ > 4 au, the density is larger than the overall trend. This is expected since at these heliocentric distances, Ulysses observes low latitude regions ($|\theta|$ < 40°, see Figure 1), which are sources of denser plasmas as shown in McComas et al. 2000.

The bottom-left panel of Figure 4 shows the corresponding 2D histogram of $\log_{10}(T)$ vs $\log_{10}(R)$. The magenta lines on the same panel show the relationship $T \propto R^{-4/3}$, which is the temperature model assuming a spherical expansion of a steady speed and adiabatic plasma with three degrees of freedom, so that $\gamma = 5/3$. Our result indicates that for most of the $R$ range we examine, the profile of the solar wind proton temperature is less steep than expected for a spherically expanding, adiabatic plasma. Instead, we observe a sub-adiabatic behavior which indicates that a heating must occur. This agrees with Richardson & Smith 2003, who construct the radial temperature profile of the protons, using Voyager observations within 1 au $< R <$ 70 au. In Section 5, we discuss possible heating mechanisms that can result to the observed proton core temperature profile. Finally, for $R < 1.38$ au and $R > 4$ au, the temperature is lower than the overall trend. As mentioned above, at these radial distances Ulysses observed plasmas originated from low-latitude regions. As shown in McComas et al. 2000, low-latitude regions are sources of colder (and denser) plasmas. Therefore, the deviation of the overall temperature profile in this $R$ range, is not related to the polytropic behavior.

Figure 4 shows a 2D histogram of $\log_{10}(T)$ vs $\log_{10}(n)$. The correlation of the two parameters confirms that, in general, the proton plasma is sub-adiabatic. With the magenta lines we show the relationship $T \propto n^{\gamma-1}$ for $\gamma = 5/3$. Clearly, the slope revealed from the observations is smaller, corresponding to $\gamma < 5/3$. Additionally, the plot suggests that the slope characterizing the density range spanning form 0.1cm$^{-3} < n <$ 1cm$^{-3}$, is smaller than the slope characterizing the data in the $n < 0.1$cm$^{-3}$ range. Finally, there is a sudden decrease in temperature for $n > 1$ cm$^{-3}$, which corresponds to the equatorial flows of slower, denser, and colder solar wind as we discuss above (see also McComas et al. 2000).

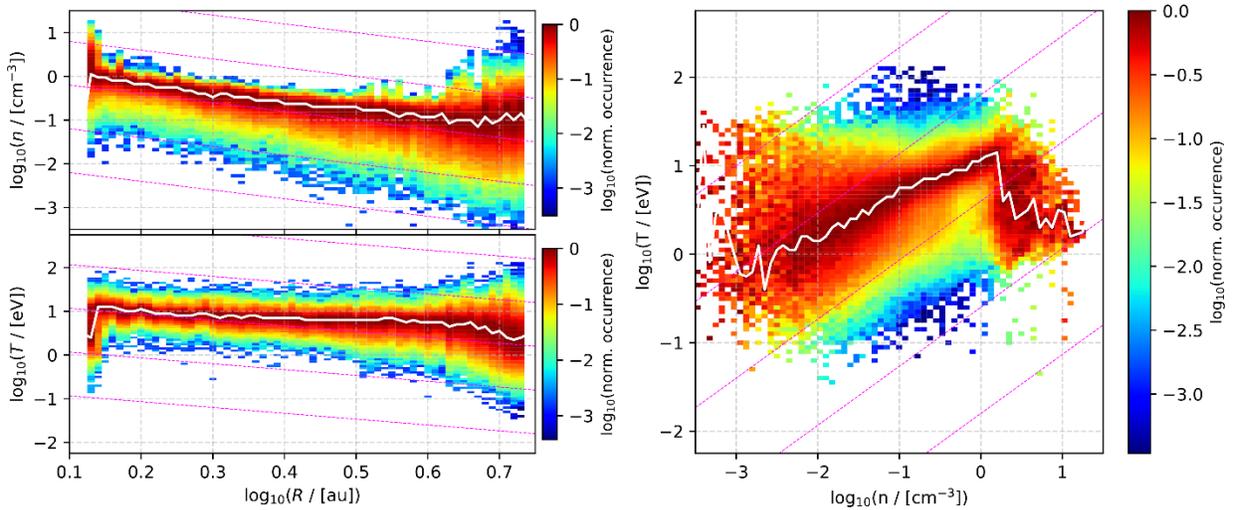

**Figure 4.** Large-scale variations of plasma density and temperature. (Top left) 2D histogram of $\log_{10}(n)$ vs $\log_{10}(R)$. The magenta lines show the $n \propto R^{-2}$ relationship. (Bottom left) 2D histogram of $\log_{10}(T)$ vs $\log_{10}(R)$. The magenta lines on the same panel show the relationship $T \propto R^{-4/3}$. (Right) 2d histograms of $\log_{10}(T)$ vs $\log_{10}(n)$, while the magenta lines on the same panel show the relationship $T \propto n^{2/3}$.

### 4.2. Short time-scale variations

#### 4.2.1. Polytropic index over solar wind speed

We first examine histograms of the polytropic index we derive from the analysis of short-time scale subintervals (eight consecutive data-points) and its correlation with the proton flow speed averaged over each subinterval. The bottom left panel of Figure 5 shows the 2D histogram of $\gamma$ and $V$, while the top left panel shows the 1D histogram of $V$ and the right panel shows the 1D histogram of $\gamma$. We can identify two peaks in the histogram of $V$. The first peak, centered at $V \sim 430$ kms$^{-1}$, corresponds to the slow solar wind observed in low latitudes ($|\Theta| < 30°$) and the second peak, centered at $V \sim 770$ kms$^{-1}$, corresponds to the fast solar wind which is observed in high latitudes ($|\Theta| > 30°$). Note however, that this structure is very different around solar maximum, when the latitudinal structure of the solar wind breaks down (for more see McComas et al. 1998; 2008). The histogram of $\gamma$ values has only one peak at $\gamma \sim 1.4$. In Figure 6, we show the 2D histogram of $\gamma$ and $V$ with each column normalized to the maximum occurrence observed in the corresponding $V$ bin. As explained in Section 3.1, after this normalization, the correlation between the two parameters is directly revealed, given that there are not extreme values in the data-set. On the same panel we show the mean value of $\gamma$ (white data-points) in each $V$ bin. According to this histogram there is not any apparent correlation between $\gamma$ and $V$. This result agrees with studies of proton plasmas at $\sim$1au (e.g., Livadiotis 2018a,b; Nicolaou & Livadiotis 2019) and in the inner heliosphere (e.g., Totten et al. 1995; Nicolaou et al 2020).

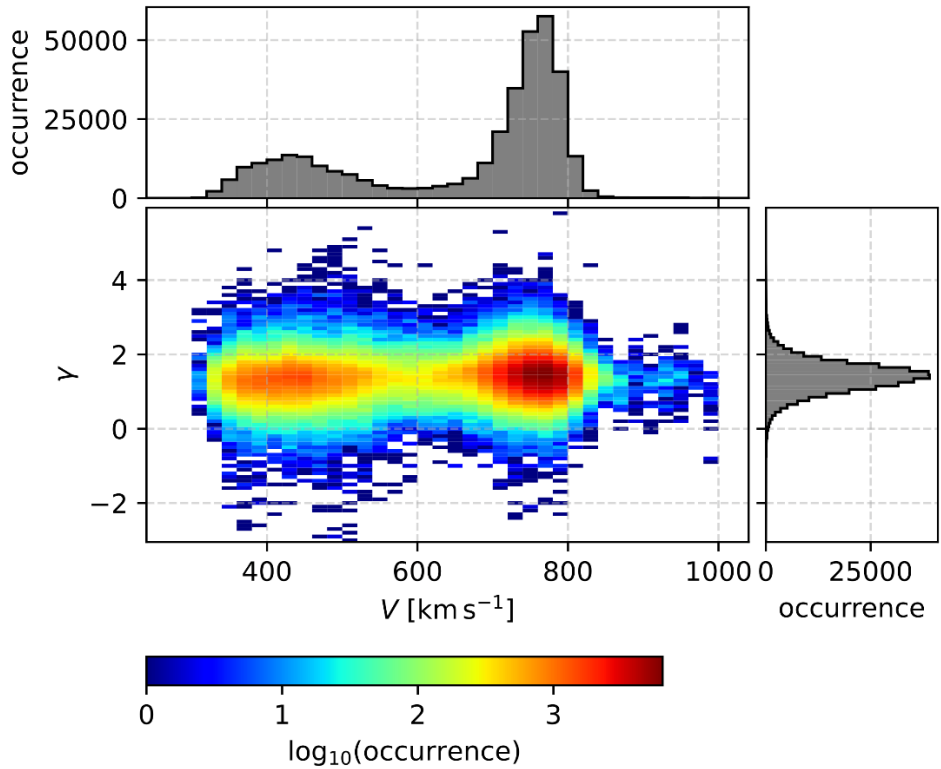

**Figure 5**. Histograms of $\gamma$ and $V$. (Bottom left) 2D histogram of the polytropic index and the solar wind proton flow speed, (top left) 1D histogram of the proton flow speed and (right) 1D histogram of the derived polytropic index.

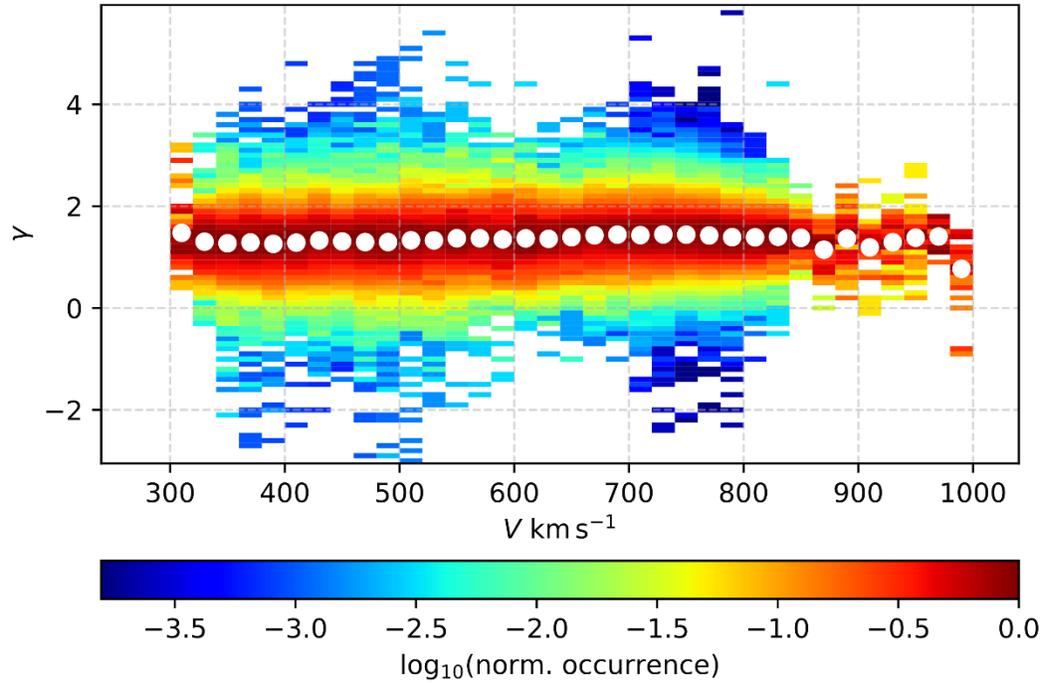

**Figure 6.** 2D histogram of $\gamma$ and $V$, with each column normalized to the maximum occurrence observed in the corresponding $V$ bin. The white data-points show the mean value of $\gamma$ in each $V$ bin. The standard error of the mean values of $\gamma$ are too small, therefore not shown here.

### 4.2.2. Polytropic index over heliocentric distance and heliographic latitude

In Figure 7, we show 2D histograms of $\gamma$ and $V$, for specific $R$ bandwidths with $\Delta R$=0.5 au. We use the same format as in Figure 6 in terms of normalizing the occurrence and calculating the average value of $\gamma$ (white data-points) and its standard error (error bars), for each $V$ bin. The range of the observed $V$ is different for different $R$ (for a different panel in Figure 7) since the solar wind speed is organized by the latitude, and the latitude observed by the spacecraft orbit is a function of the heliocentric distance. For example, the fast flows in the polar regions are observed only over heliocentric distances between ~1.75 and ~4.25 au (see also Figure 1). In general, the plots in Figure 7 confirm that there is not any apparent, systematic correlation between $\gamma$ and $V$. In Figure 8, we show the mean value of $\gamma$ and its standard error (error bars), calculated for each $R$ bin (for each panel) shown in Figure 7. According to our result, the polytropic index fluctuates between 1.38 and 1.47 at $R < 4.5$ au and drops by $\delta\gamma$~0.1 for $R > 4.5$ au. We discuss further this result in the next section.

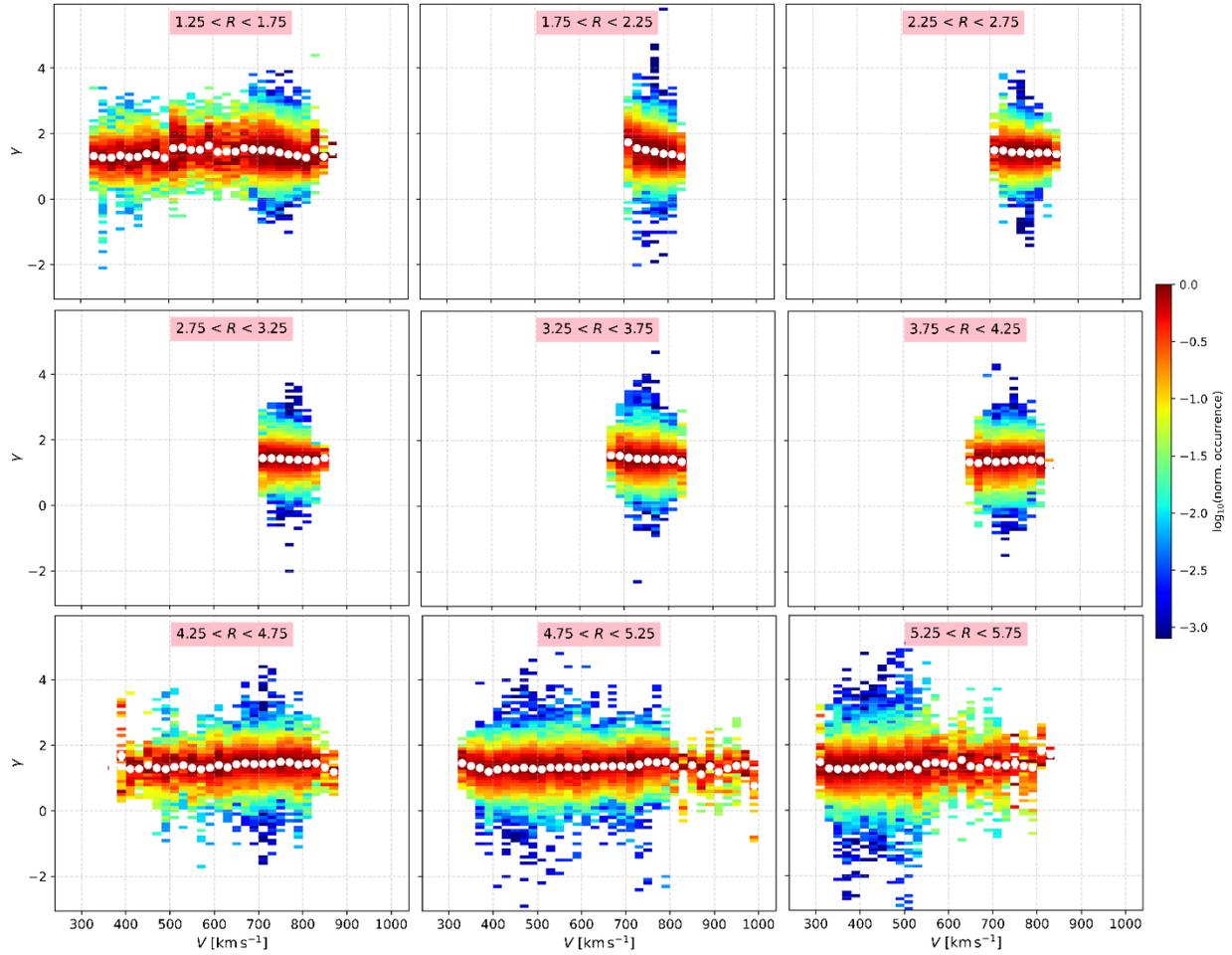

**Figure 7.** 2D histograms of $\gamma$ vs $V$ at different heliocentric distances. The occurrence in each $V$ bandwidth is normalized to the maximum occurrence value within the $V$ bin. At the top of each panel, we show the heliocentric distance bandwidth that corresponds to the plot.

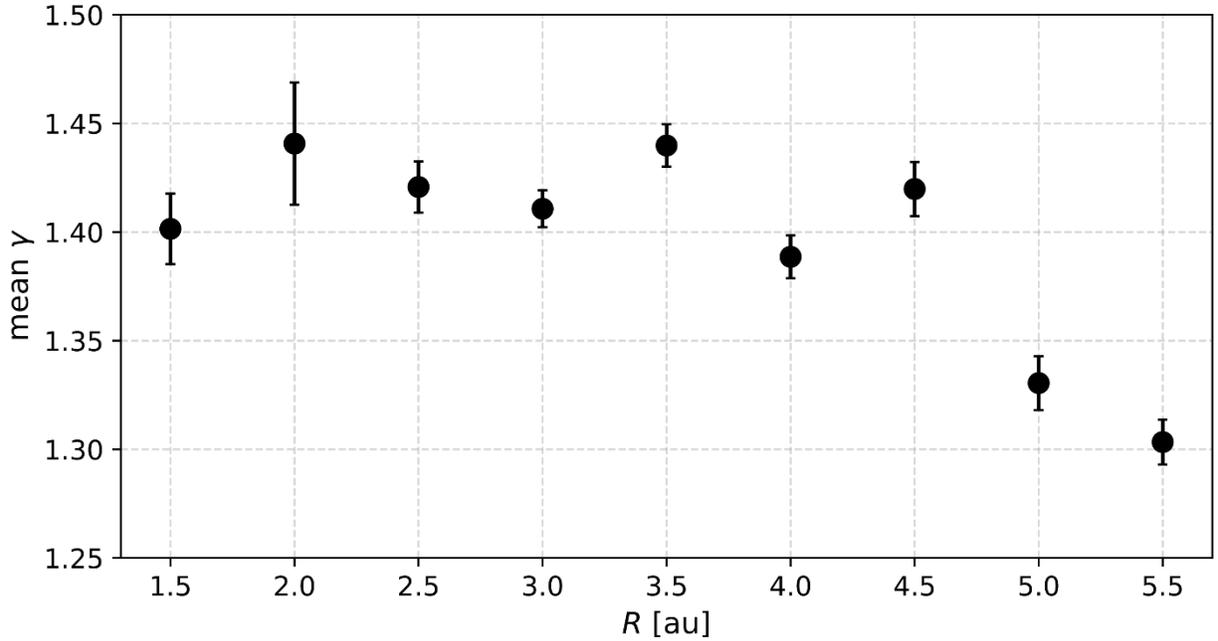

**Figure 8.** Mean value of $\gamma$ as a function of the heliocentric distance. The error bars correspond to the standard error of the mean value.

In order to examine a possible dependence of $\gamma$ on $\Theta$, we plot 2D histograms of $\gamma$ and $V$ for different $\Theta$ bins (Figure 9). Again, there is no evidence for a systematic correlation between $\gamma$ and $V$. We then calculate the mean value of $\gamma$ and its standard error for each $\Theta$ bin, which we show in Figure 10. The result indicates that $\gamma$ increases slightly towards the Sun's poles. The mean value of the polytropic index is $\gamma \pm \delta\gamma = 1.45 \pm 0.02$ at $70° < \Theta < 90°$, and $\gamma \pm \delta\gamma = 1.42 \pm 0.02$ at $-90° < \Theta < -70°$. However, $\gamma$ drops to $\gamma \pm \delta\gamma = 1.29 \pm 0.01$ at $-10° < \Theta < 10°$.

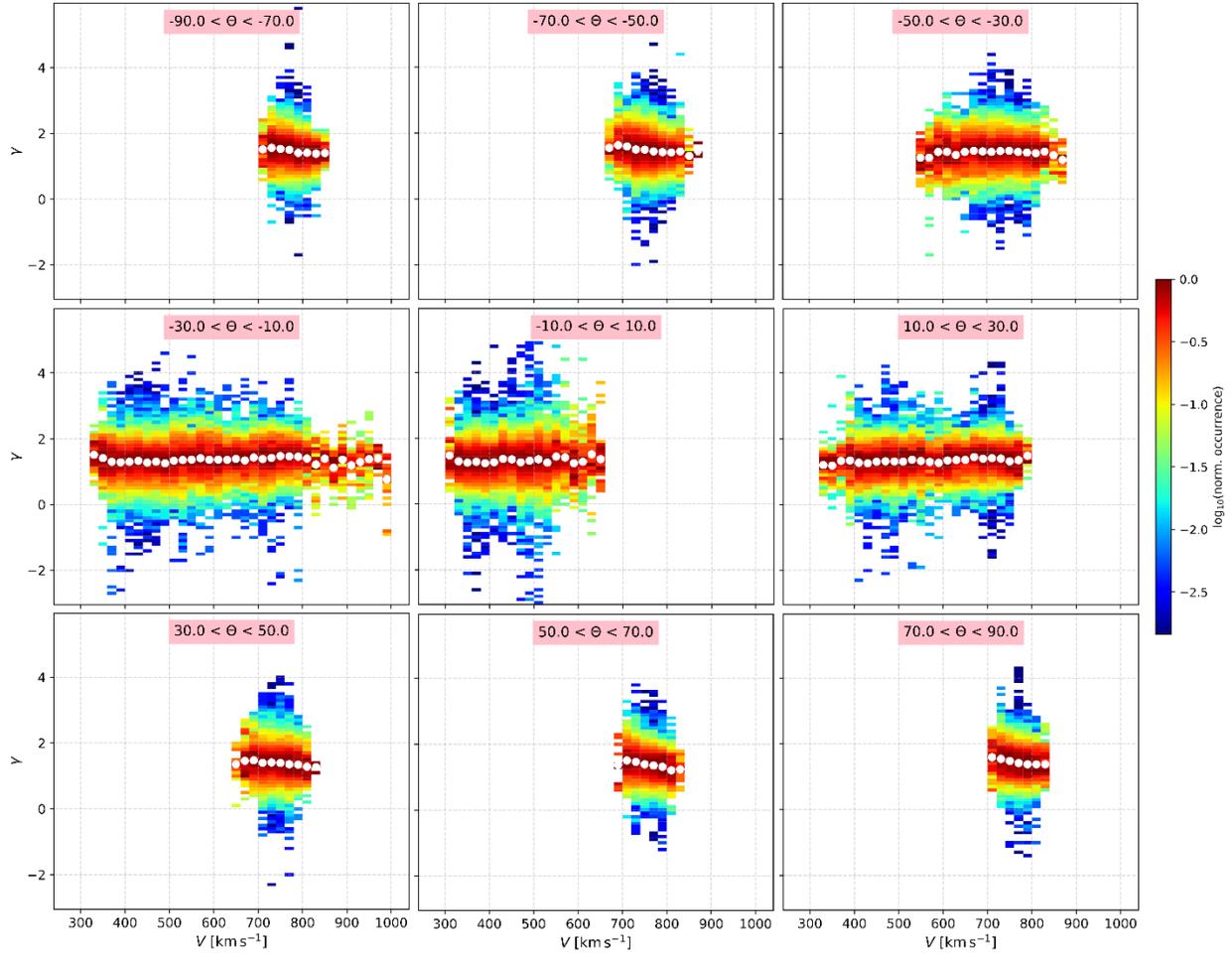

**Figure 9.** 2D histograms of $\gamma$ vs $V$ at heliographic latitudes. The occurrence in each $V$ band is normalized to the maximum occurrence value within the $V$ bin. At the top of each panel, we show the heliographic latitude band that corresponds to the plot.

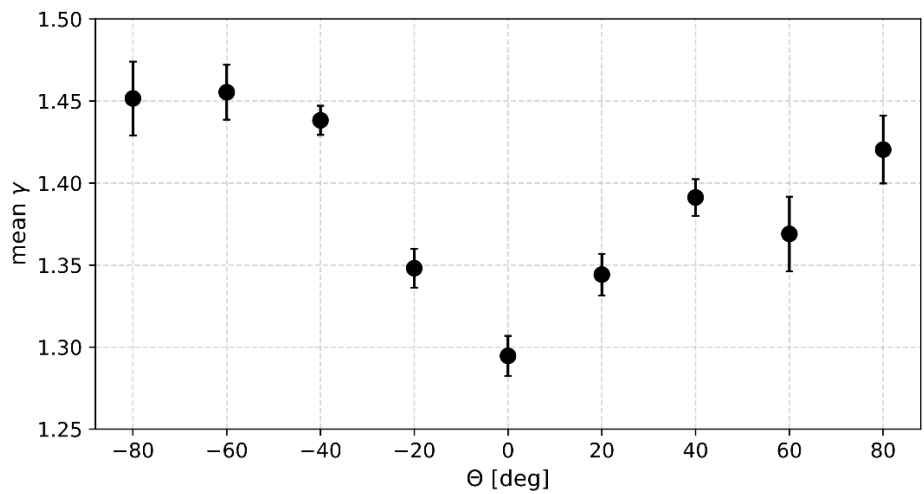

**Figure 10.** Mean value of $\gamma$ as a function of the heliographic latitude.

For completeness, in Figure 11, we show the 2D histogram of the mean $\gamma$ and its standard error $\delta\gamma$ as functions of $R$ and $\Theta$ in a polar coordinate system. For these 2D histograms we use a resolution $\delta R \times \delta\Theta$=0.25au×5°. We can clearly see that the estimated polytropic index decreases slightly with increasing $R$ and for $\Theta \rightarrow 0°$. Additionally, the plot confirms that $\gamma$ is slightly higher in the south pole than it is in the north pole. The standard error of $\gamma$ seems to be higher in the equatorial regions close to the Sun, where the spacecraft was moving fastest and spent the least time, and in mid, positive latitudes away from the Sun.

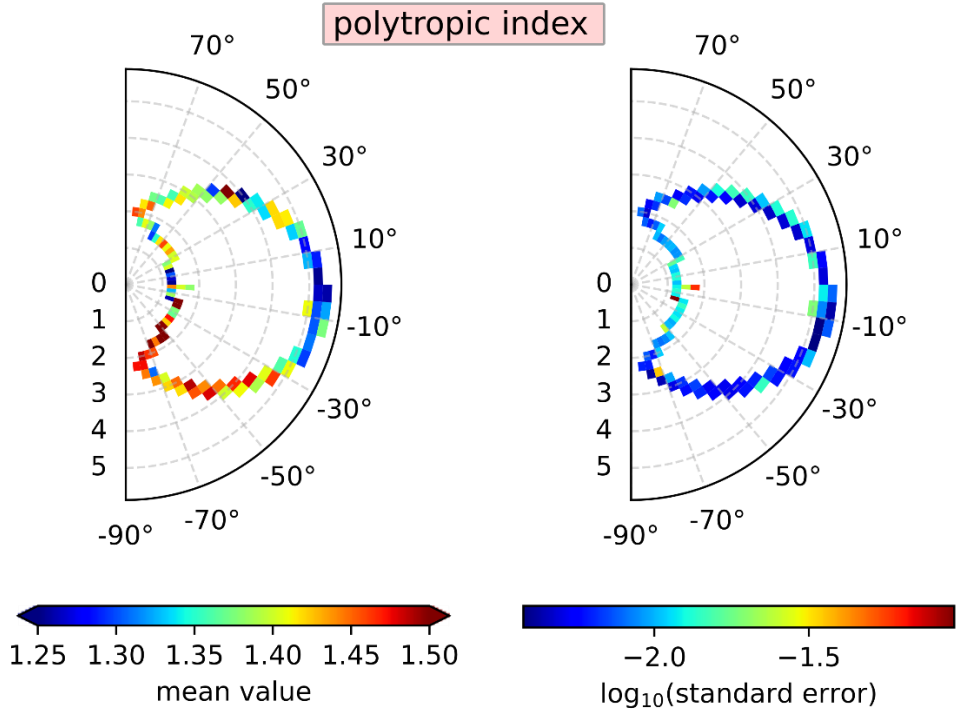

**Figure 11.** 2D histograms of (left) $\gamma$ and (right) its standard error as functions of the heliocentric distance and heliographic latitude.

### 4.2.3 Homogeneity and polytropicity

The polytropic relationship should be determined in analyses of observations of homogeneous plasmas (e.g., Newbury et al. 1997). In other words, the polytropic relationship is applicable in plasmas within the same streamline (e.g., Totten et al., 1995; Kartalev et al. 2006; Nicolaou et al., 2014b; Nicolaou & Livadiotis 2017). As we explain in Section 3.2, we use a simplified method to quantify the homogeneity of the analyzed subintervals through the Bernoulli integral variation, which is based on the method proposed by Kartalev et al. 2006. In the left panel of Figure 12, we show the 2D histogram of $\frac{\text{std}(BI)}{\text{mean}(BI)}$ vs $R$ and $\Theta$ in the same format as in Figure 11. According to this plot, the relative variation of the Bernoulli integral is < 4% for the entire data-set we use and decreases with increasing radial distance. The middle and the right panel of Figure 12 show 2D histograms of the Pearson correlation coefficient and the residuals of the linear fitting we apply to the data within the analyzed subintervals, respectively, in the same format as in Figure 11. Both magnitudes are relevant to the "goodness" of the linear fit of Eq. 1 to the data-points within each subinterval. Both plots support that the polytropic model describes better the observations obtained closer to the Sun. In Section 5, we discuss these results in much more detail.

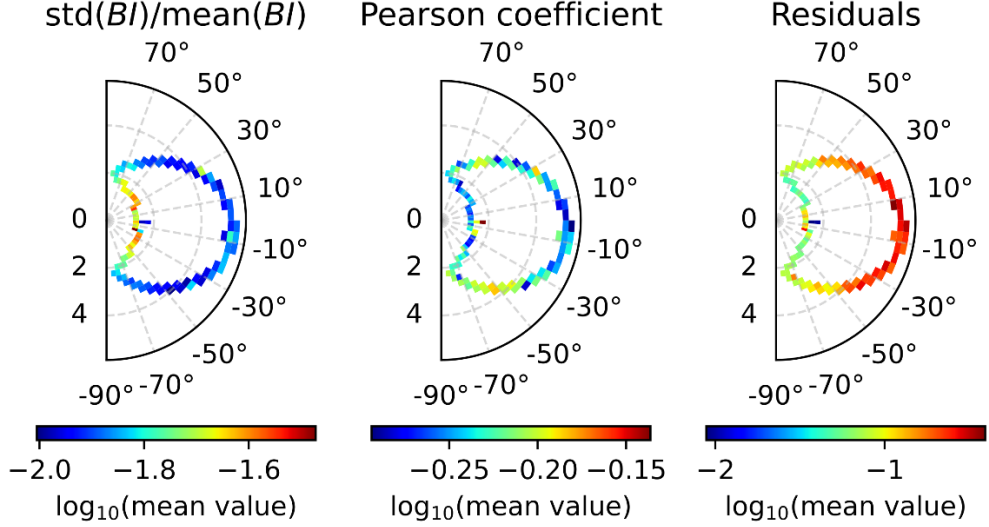

**Figure 12.** (Left) The relative deviation of the Bernoulli integral within the subintervals we analyze to derive the polytropic index, as function of the heliocentric distance and the heliographic latitude. (Middle) Pearson correlation coefficient, and (right) fitting residuals between the polytropic model and the data within the analyzed subintervals, both as functions of the heliocentric distance and the heliographic latitude.

### 4.2.4 Resolution of bias caused by density errors

Nicolaou et al., 2019 showed that typical linear fits of Eq. 2 to spacecraft observations with statistical uncertainties, lead to systematic errors in the calculation of $\gamma$. More specifically, statistical uncertainties in the plasma density result in calculations of $\gamma$, artificially closer to the isothermal value ($\gamma = 1$). On the other hand, the determination of the special polytropic index $\nu \equiv \frac{1}{\gamma-1}$, by fitting $\log_{10}(n)$ vs $\log_{10}(T)$ with:

$$\log_{10}(n) = \frac{1}{\gamma-1}\log_{10}(T) + \text{const} = \nu\log_{10}(T) + \text{const}, \tag{4}$$

suffers systematic errors ($\nu$ shifting towards 0), depending on the statistical uncertainties of the plasma temperature. The authors propose a method which excludes biased derivations, by selecting only the intervals for which the determined $\gamma$ and $\nu$ satisfy:

$$\left|\gamma - \frac{1}{\nu} - 1\right| < a, \tag{5}$$

where $a$ is the desired accuracy threshold.

In this section, we would like to verify the discovered trends of $\gamma$, using only the intervals that satisfy the criterion in Eq.5. We adopt the suggested data-filtering, using a threshold $a = 0.25$. Figure 13 shows the determined $\nu$ as a function of the determined $\gamma$. The dashed line shows the definition $\nu \equiv \frac{1}{\gamma-1}$. The blue data-points correspond to the derivations that pass the applied filter, while the red data-points correspond to derivations that do not.

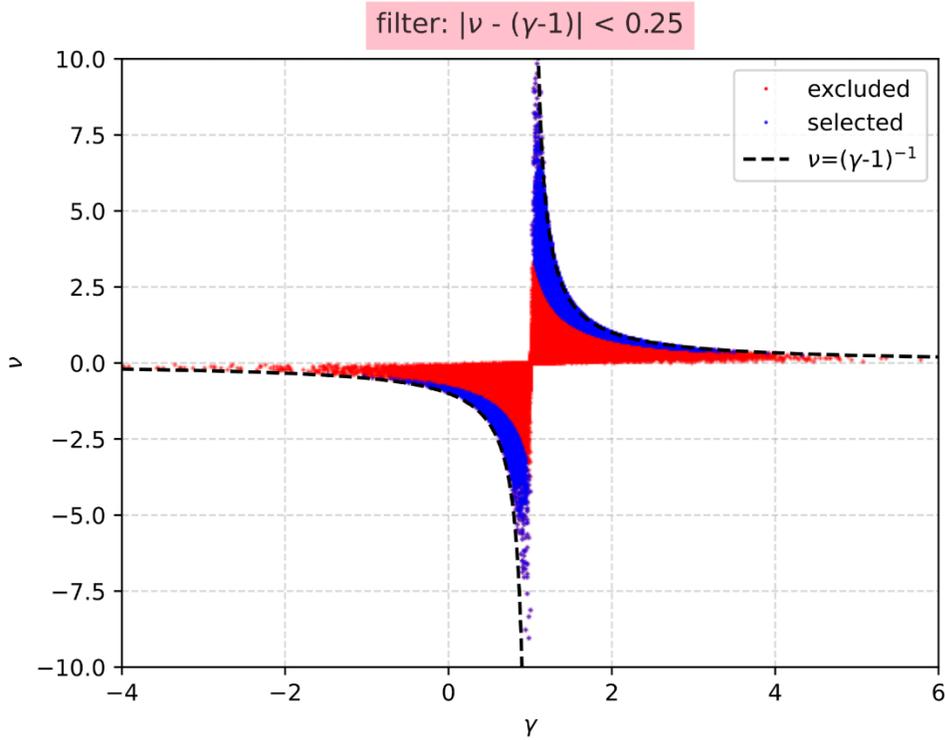

**Figure 13.** The special polytropic index $\nu$ as a function of the polytropic index $\gamma$, determined from linear fits to $\log_{10}(n)$ vs $\log_{10}(T)$ and fitting $\log_{10}(T)$ vs $\log_{10}(n)$ observations, respectively. The dashed line shows the definition $\nu \equiv \frac{1}{\gamma-1}$. We apply the filter suggested by Nicolaou et al. 2019, be selecting only data-points which are within 0.25 of the definition (blue data-points). The red-data points are excluded for this test.

In Figure 14 (a), we show the 2D histogram of $\gamma$ and $V$ using the filtered $\gamma$ values and using the same format as in the bottom left panel of Figure 5. The filter removes a significant number of isothermal values, resulting in the deep of the histogram at $\gamma = 1$. Nevertheless, there is nothing different regarding the most frequent value of $\gamma$ and its independence on $V$. In Figures 14 (b) and (c), we show the mean $\gamma$ as a function of $R$ and $\Theta$, respectively. We use the same method as explained in Section 3, but we use only the values that satisfy the applied filter. The profiles of $\gamma$ verify the behavior of $\gamma$ shown in Figures 8 and 10. The mean values in Figure 14 however, are slightly larger, due to the removal of the isothermal values.

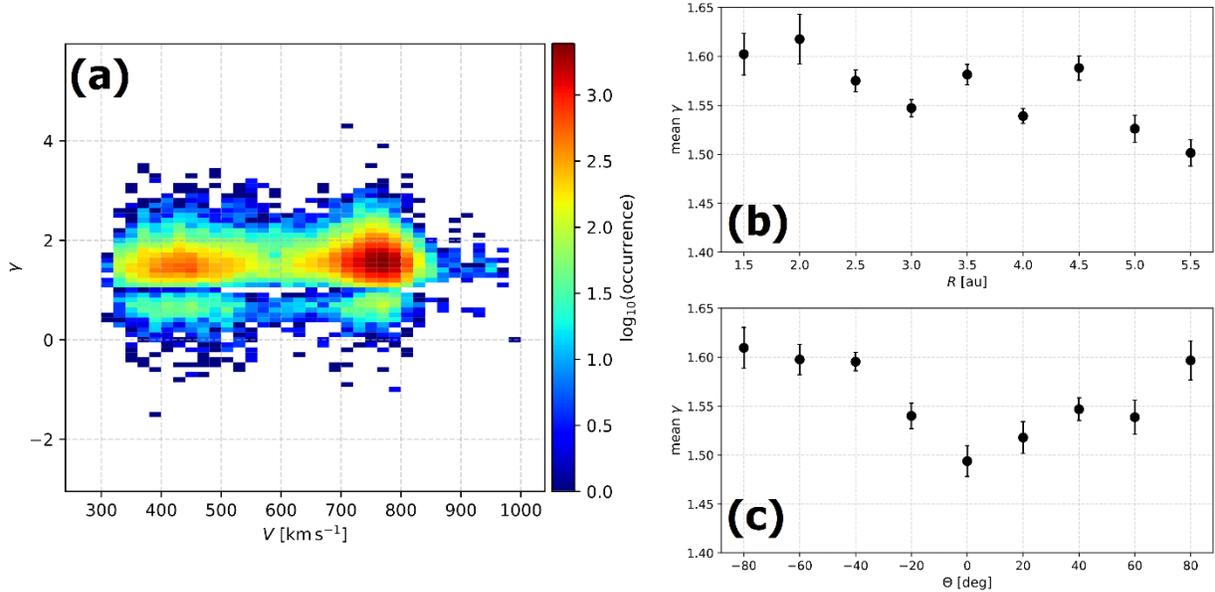

**Figure 14.** The main results of this study, using only the calculations from the filtered intervals. (a) 2D histogram of the polytropic index and proton bulk speed, (b) the mean polytropic index as a function of the radial distance, and (c) as a function of heliolatidute (see text for details).

## 5. Discussion-Conclusions

Observations of solar wind protons, obtained by Ulysses during solar minimum and within the heliocentric distance range from ~1.5 to ~5.5 au, support the conclusion that the thermal proton plasma is sub-adiabatic. The polytropic behavior is evident in both large-scale and small-scale variations of the proton density and temperature. The large-scale variations in general, agree with previous studies in the inner and outer heliosphere. Totten et al. 1995, determined $\gamma \sim 1.46$ for solar wind protons, by characterizing the radial profiles of proton density and temperature observed by Helios between 0.3 and 1 au. Huang et al. 2020 and Nicolaou et al 2020, used observations by Parker Solar Probe between 0.27 and 1 au and showed that a sub-adiabatic model describes the temperature profile reasonably well. Moreover, studies of the thermal (core) proton temperature in the outer heliosphere using observations by Voyager 2 and New Horizons (e.g., Richardson & Smith 2003, Zank et al. 2018) show additional evidence that thermal protons absorb energy as they propagate outwards. Richardson & Smith 2003 suggest that thermal protons in the inner heliosphere gain energy (are heated) from stream interaction and/or shocks. In the outer heliosphere, the heating is supplied by pick-up ions. Zank et al. 2018 extends the classical models by Holzer 1972 and Isenberg 1986 to develop a general theoretical model which describes the heating of solar wind thermal protons by low-frequency turbulence, excited by pick-up ions. The radial profiles of the solar wind proton parameters produced by the model are in excellent agreement with the observations.

We attempt to derive $\gamma$ for individual plasma streamlines by investigating the small-scale fluctuations of $n$ and $T$. Our small-scale fluctuation analysis determines an average polytropic index $\gamma \sim 1.4$. This is a smaller $\gamma$ than the one calculated from observations at ~1au in previous studies. For instance, Nicolaou et el. 2014b use observations from multiple spacecraft at ~1au and calculate an average proton plasma polytropic index $\gamma \sim 1.8$. Several other studies calculate a nearly adiabatic proton plasma at ~1 au (e.g.,

Livadiotis 2018a,b, 2019; Nicolaou & Livadiotis 2019). The more recent study by Nicolaou et al., 2020 uses observations by the Faraday Cup on board Parker Solar Probe and shows that although the large-scale variations of the plasma density and temperature reveal a nearly adiabatic plasma, the short scale variations are characterized with an average $\gamma \sim 2.7$. On the other hand, Elliott et al., 2019 analyze New Horizons observations in the outer heliosphere and show that the proton plasma parameters follow a polytropic behavior with $\gamma \sim 1.3$ at ~20 au, and $\gamma$ decreases with radial distance such that $\gamma < 1$ for $R > 30$ au. Livadiotis 2019 combines these results with the results of different studies in the inner and the outer heliosphere, to support that the polytropic index decreases gradually from $\gamma \sim 5/3$ at $R \sim 1$ au to $\gamma \sim 0$ in the inner heliosheath (e.g., Livadiotis et al. 2013; Livadiotis & McComas 2013).

Our study adds to our knowledge of the polytropic index value and its variations in the $R$ range from ~1 to 6 au, and over the heliographic latitude. According to our Figure 8, $\gamma$ decreases by $\delta\gamma \sim 0.1$ as the heliocentric distance increases from 4.5 au to 5.5 au. In order to increase the statistical accuracy of the data-points, in Figure 8, we had to use relatively large $R$ bins ($\Delta R = 0.5$ au). However, the same dependence of $\gamma$ on $R$ is supported by Figure 11 in which we use smaller $R$ bins ($\Delta R = 0.25$ au) and thus, we have more samples of $\gamma$ over $R$. Since we do not see a drastic change of $\gamma$ over the range of $R$ we examine here, we cannot provide accurate estimations of the radial profile of $\gamma$. But, if the polytropic index decrement with $R$ is real, it suggests that heat is more effectively provided to the plasma in larger radial distances, or that the plasma particle effective degrees of freedom increase as the plasma flows outwards. Considering that the ionization cavity of hydrogen is estimated to be within ~ 4 - ~5 au (McComas et al. 1999) and taking into account the radial profile of the large-scale solar wind temperature, it is possible that ions beyond 4-5 au are heated more effectively by low-frequency turbulent which is triggered by pick-up ions.

Our study indicates that the proton polytropic index is varying with the heliographic latitude as well. More specifically, the mean value of $\gamma$ decreases from ~1.45 to ~1.3 as $\Theta$ changes from -90° to 0°. Then, the mean $\gamma$ increases from ~1.3 to ~1.4 as $\Theta$ increases from 0° to 90°, creating a slight asymmetry in the $\gamma$ vs $\Theta$ profile. It is possible that the observed change in the polytropic index over $\Theta$ reflects the different thermodynamic properties of solar wind plasma originated in different solar regions. More specifically, our results support that the "coronal holes" plasma is more adiabatic that the plasma in the equatorial region. This would mean that, if we assume spherical expansion, the temperature of the plasma in the equatorial region drops less effectively than the temperature of the plasma in the polar regions. If we adopt the hypothesis of protons being heated by the dissipation of low-frequency waves generated by pick-up ions, then, the heating is not uniformly supplied, at least within the examine $R$ range. A non-uniform shape of the ionization cavity could be the reason for this.

We have also calculated the Bernoulli integral of the plasma protons, assuming a nearly adiabatic plasma. According to Nicolaou et al., 2021, this is a valid assumption of the solar wind at ~1 au. The same assumption holds in our case, since the average $\gamma$ is 1.4, and the solar wind flow speed is almost always ten times (or more) larger than the thermal speed (see Appendix A). The relative variation of the Bernoulli integral over the examined intervals is a measure of the plasma homogeneity. According to Figure 12, the homogeneity of the plasma protons increases with increasing $R$ and $\Theta$. This could mean that the plasma originated in the polar region (fast solar wind from the polar coronal holes) is less turbulent than the plasma originated in the equatorial latitudes (slow solar wind), and that the plasma becomes more homogeneous as it propagates outwards.

In the middle panel of Figure 12, we show the correlation between the data-points and the polytropic model we fit to them, while the right panel of the same figure shows the residuals between the data-points and the model, as functions of $R$ and $\Theta$. Our analysis supports the notion that the linear model fits best the observations obtained closer to the Sun, even though the Bernoulli integral has larger relative fluctuations there. Thus, this result could be due to larger statistical fluctuations of $n$ and $T$ in larger $R$.

Finally, we repeat the core analysis using the data-filtering suggested by Nicolaou et al. 2019 and show that even after this strict data-filtering, we determine the same trends for $\gamma$ vs $R$ and $\gamma$ vs $\Theta$. By doing this, we eliminate the possibility that our results suffer the systematic biases caused by statistical errors in density measurements, as predicted by Nicolaou et al. 2019.

Acknowledgements

G.L. and D.J.M were supported in part by the IMAP mission (80GSFC19C0027) as a part of NASA's Solar Terrestrial Probes (STP) mission line.

**APPENDIX A: BERNOULLI INTEGRAL**

In Figure A, we show the Bernoulli integral and its three terms (dynamic, thermal, and magnetic) using Ulysses observations from 01/01/1992 to 31/12/1998 and assuming $\gamma=5/3$ (Eq.). The thermal and the magnetic terms are considerably smaller (at least ten times) than the dynamic term of the integral for most of the time we examine here. Similar to the Bernoulli integral of protons at ~1au examined by Nicolaou et al 2021a, the thermal term does not become dominant for the range of $\gamma$ we estimate in our study. Therefore, the estimations of the Bernoulli integral do not suffer significant errors by assuming an adiabatic plasma in our case. In a more simplified approach, the integral could be estimated by the calculation of the dynamic term only (for more see Nicolaou et al. 2021a).

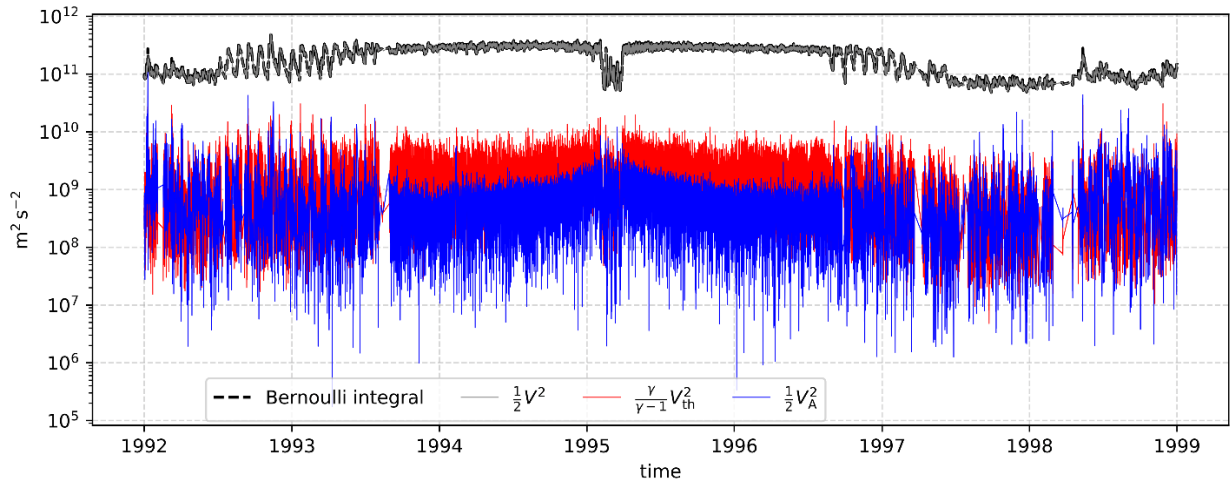

**Figure A1.** The Bernoulli integral and its terms for the time interval we analyze in this study. The value of the integral (dashed black) is almost identical with the dynamic term (grey) which is the dominant term. The thermal term (red) which assumes adiabatic plasma ($\gamma=5/3$) and the magnetic term (blue) are by more than one order of magnitude smaller than the dominant term for the most time.

**References**

Arridge, C.S., McAndrews, H.J., Jackman, C.M., et al. 2009, P&SS, 57, 2032

Bame, S.J., McComas, D.J., Barraclough, B. L., et al. 1992, A&Ass, 92, 237

Bavassano, B., Bruno, R., Rosenbauer, H. 1996, Annal. Geophys., 14, 510

Chandrasekhar S. 1967 An Introduction to the Study of Stellar Structure (New York: Dover Publications)

Coburn, J. T., Chen, & C. H. K., Squire, J. 2022, arXiv:2203.12911 [physics.space-ph]

Dialynas, K., Roussos, E., Regoli, L., et al. 2018, JGR, 123, 8066

Elliott, H. A., McComas, D.J., Zirnstein, E. J., et al. 2019, ApJ, 885, 156

Holzer, T. E. 1972, JGR, 77, 5407

Isenberg, P. A. 1986, JGR, 91, 9965

Huang, J., Kasper, J. C., Vech, D., et al. 2020, ApJS, 246, 70

Kartalev, M., Dryer, M., Grigorov, K., et al. 2006, JGR, 111, A10107

Kuhn, S., Kamran, M., Jelić, N., et al. 2010, AIP Conference Proceedings, 1306, 216

Livadiotis, G. 2016, ApJS, 223, 13

Livadiotis, G. 2018a, JGR, 123, 1050

Livadiotis, G. 2018b, Entr, 20, 799

Livadiotis, G. 2019, Entr, 21, 1041

Livadiotis, G., & Desai, M. I. 2016, ApJ, 829, 88

Livadiotis, G., & Nicolaou, G. 2021, ApJ, 909, 127

Livadiotis, G., & McComas, D. J. 2013, JGR, 118, 2863

Livadiotis, G., McComas, D. J., Dayeh, M. A., et al. 2011, ApJ, 734, 1

Livadiotis, G., McComas, D. J., Schwadron, N. A., et al. 2013, ApJ, 762, 134

McComas, D. J., Barraclough, B. L., Funsten, H. O., et al. 2000, JGR, 105( A5), 10419

McComas, D. J., Bame, S. J., Barraclough, B. L., et al. 1998, GRL, 25, 1

McComas, D. J., Ebert, R. W., Elliott, H. A., et al. 2008, GRL, 35, L18103

McComas, D. J., Funsten, H. O., Gosling, J. T., Pryor, W. R. 1999, GRL, 26, 2701

Newbury, J. A., Russel, C.T., & Lindsay, G. M. 1997, GRL, 24, 1431

Nicolaou, G., Livadiotis, G., Wicks, R.T., et al. 2020, ApJ, 901, 26

Nicolaou, G., & Livadiotis, G. 2017, ApJ, 838, 7

Nicolaou, G., & Livadiotis, G. 2019, ApJ, 884, 52

Nicolaou, G., Livadiotis, G., & Moussas, X.D. 2014b, SolPhy, 289, 1371

Nicolaou, G., Livadiotis, G., & Wicks, R.T. 2019, Entr, 21, 997

Nicolaou, G., McComas, D. J., Bagenal, F., Elliott, H. A. 2014a, JGR, 119, 3463

Nicolaou, G., McComas, D. J., Bagenal, F., et al. 2015, P&SS, 111, 116

Nicolaou, G., Livadiotis, G., & Desai, M. I. 2021a, Appl. Sci., 11, 4643

Nicolaou, G., Livadiotis, G., & Desai, M.I. 2021b, Appl. Sci., 11, 4019

Nicolaou, G., Wicks, R. T., Owen, C. J., et al. 2021c, A&A, 656, A10

Pang, X., Cao, J., Liu, W., et al. 2015, JGR, 120, 4736

Park, J.-S., Shue, J.-H., Nariyuki Y., et al. 2019, JGRA, 124, 1866

Richardson, J. D., & Smith, C. W. 2003, GRL, 30, 1206

Totten, T. L., Freeman, J. W., & Arya, S. 1995, JGR, 100, 13

Verscharen, D., Klein, K.G. & Maruca, B.A. 2019, LRSP, 16, 5

Wu, H., Verscharen, D., Wicks, R.T., et al. 2019, ApJ, 870, 106

Zank, G.P., Adhikari, L., Zhao, L.-L., et al 2018 ApJ 869 23